\documentclass[10pt,prd,twocolumn,showpacs,amssymb,superscriptaddress,aps,nofootinbib,noeprint,nolongbibliography
]{revtex4-2}

\usepackage{graphicx,amsmath,amssymb}
\usepackage{bm}
\usepackage[colorlinks=true]{hyperref}
\hypersetup{colorlinks=true,
            citecolor=blue,
            linkcolor=red,
            urlcolor=red}

\usepackage{amssymb,amsmath,amsfonts,mathrsfs,url}
\usepackage{xcolor}
\usepackage[colorlinks=true]{hyperref}
\hypersetup{citecolor=blue,linkcolor=red}
\usepackage{multirow,array}
\DeclareMathAlphabet{\pazocal}{OMS}{zplm}{m}{n}

\newcommand{\rH}{r_{\mathcal{H}}}

\newcommand{\aD}{a_{\rm D}}

\usepackage{pifont}% http://ctan.org/pkg/pifont
\newcommand{\cmark}{\ding{51}}%
\newcommand{\xmark}{\ding{55}}%

\newcommand{\boldl}{\boldsymbol{\ell}}
\newcommand{\boldn}{\boldsymbol{n}}
\newcommand{\boldm}{\boldsymbol{m}}
\newcommand{\boldmb}{\boldsymbol{\bar{m}}}

\begin{document}
%%%%%%%%%%%%%%%%%%%%%%%%%%%%%%%%%%
\title{Doubly regular black holes}
\date{\today}
\author{Arthur G. Suvorov}\thanks{arthur.suvorov@tat.uni-tuebingen.de}
\affiliation{Departament de F{\'i}sica Aplicada, Universitat d'Alacant, Ap. Correus 99, E-03080 Alacant, Spain}
\affiliation{Theoretical Astrophysics, IAAT, University of T{\"u}bingen, T{\"u}bingen, D-72076, Germany}
\author{Pedro Bargueño}%\thanks{pedro.bargueno@ua.es}
\affiliation{Departament de F{\'i}sica Aplicada, Universitat d'Alacant, Ap. Correus 99, E-03080 Alacant, Spain}

%%%%%%%%%%%%%%%%%%%%%%%%%%%%%%%%%%
\begin{abstract}
\noindent In addition to curvature singularities, electrovacuum black holes in general relativity exhibit thermodynamic singularities. These so-called Davies’ points occur at nonextremal values of charge and spin where the heat capacity diverges and may indicate a type of theoretical incompleteness. The thermodynamic regularity of several families of static, asymptotically flat spacetimes with bounded curvature invariants is examined using a theory-agnostic framework, showing that, while they may be regular in physical space, they are generally not in phase space. The inclusion of angular momentum, via the Newman-Janis algorithm, makes the set of such ``doubly regular'' objects especially restrictive. It is argued that, if thermodynamic regularity is to be considered a desirable property for an astrophysical black hole, these considerations could be used to narrow down the viable pool of regular extensions to the Kerr-Newman metric.
\end{abstract}

\maketitle

%%%%%%%%%%%%%%%%%%%%%%%%%%%%%%%%%%%%%%%%%%%%%%%%%%
\section{Introduction} \label{sec:intro}
%%%%%%%%%%%%%%%%%%%%%%%%%%%%%%%%%%%%%%%%%%%%%%%%%%

Black holes (BHs) are a fundamental prediction of general relativity (GR). The celebrated ``no-hair'' relations establish the uniqueness of the Kerr-Newman family as a descriptor for these stable ultimates in asymptotically-flat electrovacua \cite{heus96}, and even in some astrophysical environments \cite{gurl15}. Aside from the inescapability of geodesic incompleteness (at least for $r \geq 0$) due to the Hawking-Penrose theorems -- a fact often used to motivate modified gravitational frameworks -- Kerr-Newman objects also suffer from \emph{thermodynamic} blowup \cite{davies78}.

Spacetime structure and thermodynamics appear to be deeply related. For example, Padmanabhan \cite{pad02} showed how one may recast the Einstein equations in a form resembling the first law of thermodynamics near a horizon, while Jacobson \cite{jacob95} has argued that the Einstein equations themselves emerge naturally if horizon entropy and area are equal up to the Hawking temperature. The latter quantities can be derived from purely geometric means using diffeomorphism invariance and Noether's theorem assuming a stationary spacetime \cite{wald93}. In direct analogy with the classical theory of thermodynamics, one can thus define a heat capacity and show that, for the Kerr-Newman family, there are singular (``Davies'') points \cite{davies78}. 

While astrophysical charge is expected to be small because a negatively (positively) charged hole will preferentially capture protons (electrons), even Kerr objects harbour Davies' points at the otherwise unremarkable spin value of $\aD = \sqrt{3 - 2 \sqrt{3}} \approx 0.681$. As angular momentum accretion is subject to no such neutralization \cite{thorne74} -- and in fact super-Davies spins have been inferred from X-ray \cite{rey19,rey21} and gravitational-wave \cite{lasky24,meh25} data, and are achieved in binary-merger \cite{eti09,rezz09} or core-collapse \cite{shib02,shib05} simulations -- this implies that astrophysical BHs may undergo a kind of thermodynamic phase transition. How such a transition may manifest is, however, unclear: there is no evidence for such a bifurcation in the sense that super-Davies objects do not seem to exhibit unusual behaviour relative to their slower counterparts. For instance, the two microquasars XTE J1550--564 and 4U 1543--47 are observationally similar -- both hosting BHs of mass $\approx 10 M_{\odot}$ and showing super-Eddington flares \cite{sob00,jin24} -- despite the former having $a \approx 0.34$ \cite{stein11} and the latter having $a \approx 0.8$ \cite{shaf06} as inferred from spectral fitting (though cf. Ref.~\cite{mcc11} for a critical discussion). 

This prompts one to question whether Davies' points are metric artifacts which may be ``regularized'', much like the physical geometry, in an ultraviolet-complete theory that eventually comes to replace GR. In this paper, we explore the nature of \emph{doubly regular} BHs: those that are free of both Davies' points and curvature singularities. Such mathematical considerations may be useful to favour certain members of the regular BH superset which are otherwise difficult to distinguish via experiments, as the regularizing hair is typically attributed to Planck-scale physics out of instrument reach. 

While a comprehensive study is made difficult by the fact that many families of regular BHs have been proposed (see Refs.~\cite{torres23,lan23} for reviews), we show that a number of popular models for static, regular BHs (e.g. Hayward \cite{hay06} and Simpson-Visser \cite{sv19}) have Davies' points. To demonstrate that double regularity can be realized though, a new hybrid-Bardeen family is introduced. On the other hand, making use of the Newman-Janis algorithm \cite{nj65} to build rotating generalizations, we show that double regularity becomes even more restrictive unless the heat capacity is positive in the $a\to0$ limit.

This short paper is organized as follows. Section~\ref{sec:bhthermo} introduces the general form of the spacetimes (Secs.~\ref{sec:static} and \ref{sec:rotating}) we consider together with relevant thermodynamic definitions (Sec.~\ref{sec:heatcapacity}) and a discussion on their physical interpretation (Sec.~\ref{sec:physical}). We define \emph{regular} in various contexts in Section~\ref{sec:notions}, which is then applied to a number of popular static families in Section~\ref{sec:almostrn}. Rotating cases are covered in Section~\ref{sec:rotfams}. We close with a summary and some discussion in Section~\ref{sec:discussion}. 

%%%%%%%%%%%%%%%%%%%%%%%%%%%%%%%%%%%%%%%%%%%%%%%%%%
\section{Black-hole thermodynamics} \label{sec:bhthermo}
%%%%%%%%%%%%%%%%%%%%%%%%%%%%%%%%%%%%%%%%%%%%%%%%%%

Here we introduce the line elements that we consider together with the thermodynamic quantities of relevance. 

%%%%%%%%%%%%%%%%%%%%%%%%%%%%%%%%%%%%%%%%%%%%%%%%%%
\subsection{Static black holes} \label{sec:static}
%%%%%%%%%%%%%%%%%%%%%%%%%%%%%%%%%%%%%%%%%%%%%%%%%%

In standard Boyer-Lindquist coordinates $\{t,r,\theta,\phi\}$ with units such that $G = c =1$, we adopt the following static, spherically-symmetric line element
\begin{equation}  \label{eq:metric}
ds^2 = -e^{2 \alpha(r)} B(r) dt^2 + \frac{dr^2}{B(r)} + r^2 d\theta^2 +
r^2 \sin^2\theta d\phi^2,
\end{equation}
where the two potentials $\alpha$ and $B$ are, in principle, to be determined from a set of field equations. As written, horizons reside at roots of the algebraic equation $0 = B(r)$; the largest of these is denoted by $\rH$---the radial location of the event horizon $\cal H$.

To introduce the Hawking temperature, it is convenient to define a complex null-tetrad basis, $\{\boldl,\boldn,\boldm,\boldmb\}$, with the overhead bar denoting complex conjugation. These vectors satisfy the pair-wise normalizations $\boldl \cdot \boldn = -\boldm \cdot \boldmb = -1$ with orthogonality between other products, and $\boldl$, evaluated at $\rH$, is such that it matches the contravariant time-like Killing vector, $\xi^{\mu} = \delta^{\mu t}\partial_{t}$, associated with the stationarity of the spacetime \eqref{eq:metric}. These conditions are enough to fix the basis, viz.
\begin{equation} \label{eq:vecs}
\begin{aligned}
    &\ell^{\mu} = \left\{1,e^{\alpha} B, 0,0\right\},\\
  &   n^{\mu} = \left\{\frac{e^{-2 \alpha}}{2B(r)},-\frac{1}{2}e^{-\alpha},0,0\right\}, \\
  & m^{\mu} = \left\{0,0, \frac{i}{\sqrt{2} r},\frac{1}{\sqrt{2} r \sin\theta}\right\}.
    \end{aligned}
\end{equation}
From \eqref{eq:vecs}, the Killing surface gravity reads \cite{heus96}
\begin{equation} \label{eq:surfgrav}
\begin{aligned}
    \kappa_{\rm \cal H} &\equiv -\left( \ell^{\mu} n^{\nu} \ell_{\nu;\mu} \right)_{\cal H}\\
    &= \left(e^{\alpha} B_{,r} \right)|_{\cal H}.
    \end{aligned}
\end{equation}
Equipped with the above, the BH (Hawking) temperature, $T$, can be defined purely in terms of theory-independent quantities as
\begin{equation} \label{eq:temp}
T = \kappa_{\cal H} /2 \pi.
\end{equation}
Expression \eqref{eq:surfgrav}, and rotating extensions as considered next, are used to introduce the heat capacity in Sec.~\ref{sec:heatcapacity}.

%%%%%%%%%%%%%%%%%%%%%%%%%%%%%%%%%%%%%%%%%%%%%%%%%%
\subsection{Stationary black holes} \label{sec:rotating}
%%%%%%%%%%%%%%%%%%%%%%%%%%%%%%%%%%%%%%%%%%%%%%%%%%

How exactly an axially-symmetric extension of the metric \eqref{eq:metric} should behave is not obvious. The main reason for this is that the spin parameter related to angular momentum, $a \equiv J/M$, could feature within the metric components in various ways depending on the theory of gravity and/or source terms under consideration. This is a well-known issue in the context of rotating stars in GR, where the exterior spacetime cannot be described by the Kerr metric because the stellar multipole moments mandate departures from the no-hair relations when applying junction conditions at the surface \cite{quev91,sterg98}. It is also not, in general, possible to reduce the metric to a Papapetrou form because the existence of harmonic coordinates, used to gauge-reduce the number of free functions \cite{pap53}, relies on the invertibility of the Ricci tensor \cite{suvm16,xie21}. This is related to circularity: because the Killing vectors associated with stationarity and axisymmetry need not satisfy transitivity (invariance under the simultaneous mappings $t \to -t$ and $\phi \to -\phi$) in arbitrary spacetimes, additional terms such as $g_{\theta \phi}$ cannot be gauged away \cite{cart69,cart70}. Up to eight free functions can feature in general, and without a specific set of field equations there is no clear path to constraining them (see Refs.~\cite{xie21,suvg23} for discussions).

One possibility for building rotating extensions of \eqref{eq:metric} involves the eponymous Newman-Janis algorithm \cite{nj65}. Aside from its simplicity, at least in many cases this method is known to preserve the geometric regularity of a static seed \cite{bamb13}. As originally introduced, the method involves rotating the null tetrads \eqref{eq:vecs} through the complex plane, projecting the metric components into a new basis, and then slicing in such a way that a real line-element is returned. The method has since been revisited without the explicit need for complexification \cite{az14}. For our purposes, and restricting attention to the case $\alpha = 0$ for simplicity, the end result is (see, e.g., Ref.~\cite{ernesto19})
\begin{equation}\label{eq:njmet}
\begin{aligned}
ds^{2} = & \left(1-\frac{2f}{\rho^{2}}\right)dt^{2}-\frac{\rho^{2}}{\Delta}dr^{2}
+\frac{4af \sin^{2}\theta}{\rho^{2}}dt d\phi \\
&-\rho^{2}d\theta^{2}-\frac{\Sigma \sin^{2}\theta}{\rho^{2}}d\phi^{2} ,
\end{aligned}
\end{equation}
with
\begin{equation}
\rho(r,\theta) = \sqrt{ r^{2} + a^{2} \cos^{2}\theta },
\end{equation}
\begin{equation}
f(r)  =  r^{2} \left[1-B(r) \right]/{2} ,
\end{equation}
\begin{equation}
\Delta(r)  = -2f(r) + a^{2}+ r^2 ,
\end{equation}
and
\begin{equation}
\Sigma(r,\theta) = \left(r^{2}+a^{2} \right)^{2}- a^{2} \Delta(r) \sin^{2} \theta,
\end{equation}
where $a=0$ returns the line element \eqref{eq:metric}.

Importantly, at least for metrics in the form \eqref{eq:njmet}, it has been shown that the method of ``anomaly cancellation'' can be used to define the surface gravity in a particularly simple way \cite{rob05}. Such gauge symmetry anomalies -- representing the breakdown of general covariance when quantizing the equations of motion -- are often considered undesirable properties of a physical theory\footnote{The matter is subtle though if currents exist, as (for example) relates to quantum-mechanical processes associated to electric fields with chiral (`Adler') anomalies that can source an electromagnetic dynamo \cite{dp24}. Considerations of analogous ``gravitomagnetic dynamos'' lie beyond the scope of this article.}. By demanding regularity conditions on blackbody fluxes from the horizon, the surface gravity can be read off as \cite{mur06}
\begin{equation} \label{eq:kapparot}
\kappa_{\cal H} = \frac{\Omega_{\cal H}}{2a} \left( \frac{\partial \Delta}{\partial r} \right)_{r = \rH},
\end{equation}
where $\Omega_{\cal H}$ is the angular velocity, $\Omega = g_{t \phi}/g_{\phi \phi}$, evaluated at the horizon. Note that, in taking thermodynamic derivatives, one should substitute $a = J/M$ and hold $J$ fixed rather than $a$ for consistency.

While there are other methods of calculating $\kappa_{\cal H}$ that extend to cases with $\alpha \neq 0$ (such as those involving tunnelling through the effective potential-barrier \cite{wil00,ma08}), expression \eqref{eq:kapparot} provides a computationally straightforward way to obtain the temperature \eqref{eq:temp}, for the Newman-Janis extension \eqref{eq:njmet}, without requiring that limits be taken for $a > 0$. These limits become difficult to handle in cases where the horizon satisfies a transcendental equation and numerical treatments are necessary (e.g., when there are exponential suppressions \cite{sv19}).

%%%%%%%%%%%%%%%%%%%%%%%%%%%%%%%%%%%%%%%%%%%%%%%%%%
\subsection{Heat capacity and Davies' points} \label{sec:heatcapacity}
%%%%%%%%%%%%%%%%%%%%%%%%%%%%%%%%%%%%%%%%%%%%%%%%%%

The heat capacity at constant $X$ (e.g. charge, angular momentum, or more general hairs) is defined as \cite{davies78}
\begin{equation} \label{eq:heatcapacity}
C_{\rm X} = T \left( \frac{\partial S}{\partial T}\right)_{\rm X},
\end{equation}
where the subscript is understood in the familiar thermodynamic sense that one holds $X$ fixed in performing the derivative and it should be emphasized that we are implicitly discussing the capacity \emph{at the horizon}, $C_{\rm X} \equiv (C_{\rm X})|_{\cal H}$. In expression \eqref{eq:heatcapacity} we have introduced an entropy function, $S$. As shown by Wald in a seminal work \cite{wald93}, $S$ can also be derived by purely geometric arguments without appealing to any specific action. This key insight is conceptually striking: associated with the timelike Killing field is a Noether form, the surface integral of which over the horizon exactly matches the Bekenstein entropy in the GR limit up to multiplicative constants \cite{bek73}. This holds because functional variations of the mass and angular momentum of the spacetime, as defined via Komar-like integrals, are related up to geometric factors regardless of the action principle. These variational relations can be recast as a first law of BH thermodynamics, from which an entropy can be nominally introduced.

Strictly speaking the Wald formula is necessary to calculate $C_{\rm X}$, which would require us to fix our attention to a specific theory or family of theories. However, in this work we are interested only in divergences of expression \eqref{eq:heatcapacity} -- the Davies' points -- and thus such a restriction is actually unnecessary. If $X$ denotes all BH hairs aside from mass, the chain rule implies that
\begin{equation} \label{eq:entropyfin}
    \left( \frac{\partial S}{\partial T}\right)_{\rm X} =    \left( \frac{\partial S}{\partial M}\right)_{\rm X} \left( \frac{\partial M}{\partial T}\right)_{\rm X} .
\end{equation}
Provided the entropy itself is well behaved -- as expected for any physically reasonable theory of gravity -- the only way for $C_{\rm X}$ to diverge is if
\begin{equation} \label{eq:divcond}
    \left(\frac{\partial T } {\partial M}\right)_{\rm X} \to 0.
\end{equation}
Points in phase space such that condition \eqref{eq:divcond} is met are referred to as \emph{Davies' points} after Ref.~\cite{davies78}.

Since the temperature can be defined in a theory-agnostic way via expression \eqref{eq:temp}, we can investigate the existence of thermodynamic singularities using only the metric potentials defining the line element \eqref{eq:metric} or its rotating extension \eqref{eq:njmet}. Combining this with the fact that geometric regularity also depends only on the particulars of the line element, one may try to reduce the pool of viable metric potentials by appealing to \emph{double regularity} in a way that is theory agnostic.

%%%%%%%%%%%%%%%%%%%%%%%%%%%%%%%%%%%%%%%%%%%%%%%%%%
\subsection{Physical interpretation of thermodynamic singularities} \label{sec:physical}
%%%%%%%%%%%%%%%%%%%%%%%%%%%%%%%%%%%%%%%%%%%%%%%%%%

We take a brief detour to explore what a thermodynamic singularity indicates. Indeed, if one wishes to argue that an absence of Davies' points is a desirable property of an astrophysical hole, some words are warranted on what this implies; for details, we refer to Ref.~\cite{bfm24}.

A negative heat capacity designates a state where the temperature increases with the emission of radiation, meaning the object is thermally unstable. This applies to self-gravitating bodies that are not self-regulating via nuclear burning \cite{lynd77} and Schwarzschild BHs in particular, where $T \propto 1/M$, as a decrease in mass-energy leads to an increase in $T$. A positive $C_{\rm X}$ instead means that $T$ should decrease as energy is lost for fixed $X$. Such an endo/enthalpic transition, corresponding to Davies' points, invites thermonuclear ignition in main-sequence stars and allows their formation in general. 

Perhaps unintuitively, it is a negative capacity that arguably represents a \emph{cosmologically stable} BH in the following sense. A positive capacity would work against the formation of supermassive objects residing at galactic centres, as their growth would be stalled if the radiative power increases as the BH enlarges: radiative losses may eventually dominate any accretion process. Such an enhancement would appear to be in conflict with James Webb and other data, revealing a large population of supermassive BHs at high redshifts of varying mass \cite{silk24}, since one may expect a natural mass ceiling and with signatures of radiative outputs around the heaviest sources. 

The existence of a Davies' point (or non-negative capacity) hints therefore that the evolution of BH can converge to end states that are observationally questionable (depending on the numerical coefficient of the power). It may be the case however that the excess hair is radiated away faster than mass-energy is lost, meaning that long-term evolution converges to the stable regime in the sense described above. For Kerr BHs, Page \cite{page76} has shown that the radiation spectrum implies that angular momentum is lost faster than mass-energy, indicating that an object that is not receiving spin donations will eventually transition to a sub-Davies' configuration. This is \emph{not} the case for Reissner-Nordstr{\"o}m objects, which may evolve towards extremality \cite{his90}. The rate at which generalized hairs decay relative to mass requires detailed calculations in a fixed theory, lying beyond the scope of this article. Nevertheless, such issues can be avoided if the object is thermodynamically regular as no astrophysical manufacturing can trigger an enthalpic transition.

%%%%%%%%%%%%%%%%%%%%%%%%%%%%%%%%%%%%%%%%%%%%%%%%%%%%%%%%%%%%%%%%%%%%%%%%%%%
\section{Notions of regularity} \label{sec:notions}
%%%%%%%%%%%%%%%%%%%%%%%%%%%%%%%%%%%%%%%%%%%%%%%%%%%%%%%%%%%%%%%%%%%%%%%%%%%

In order to examine regularity and/or double-regularity, we need to define precisely what is meant by the word ``regular'' in various contexts.

\subsection{Geometric regularity}

The notion of geometric regularity is subtle. It is well-known that curvature invariants (contractions of the Riemann tensor and its various decompositional submembers) may be $C^{\infty}$ over the entire manifold and yet the spacetime is geodesically incomplete \cite{mac15}. 

{A complete set of Riemann invariants were introduced by Zakhary and McIntosh (ZM) \cite{zm97}. It has been shown that some BHs where the ZM invariants are finite -- what might be called ``ZM regularity'' or a \textbf{ZMR} spacetime -- are not geodesically complete (\textbf{GC}) \cite{zm23,mag24}. The results in Ref.~\cite{zm23} include certain versions of the Simpson-Visser metric, for instance. For simplicity, we consider the definition}
\begin{itemize}
\item[]{\textbf{Curvature regularity (CR):} $\mathcal{K} \equiv R_{\mu \nu \alpha \beta} R^{\mu \nu \alpha \beta}$ is bounded over physical space.}
\end{itemize}
{It should be understood that one might have $\textbf{CR} \subset \textbf{GC}$ in some appropriate sense of the categories (cf. Refs.~\cite{beem76,coley09}). Furthermore, not \emph{all} of the ZM invariants need to be bounded for what we call $\textbf{CR}$ to hold (see Ref.~\cite{pz05} for constructions involving `truly naked BHs').}

{At a geometric level, other types of singular behaviour can arise due to a mass-inflation instability associated with a Cauchy horizon. The Reissner-Nordstr{\"o}m metric exhibits such a pathology as infalling radiation approaching the Cauchy horizon suffers an infinite blueshift \cite{pi90}, though inclusion of a positive cosmological constant may tame the blowup \cite{card18}. In analogy with our previous definitions, this invites another notion of regularity that could perhaps be called ``Cauchy-horizon regular'' (\textbf{CHR}). Very recently, \citet{bon25} have shown that \textbf{CR} BHs with inner horizons may be immune to this instability because the relevant mass function grows only polynomially rather than exponentially. There seems support then for the conjecture that $\textbf{CHR} \subset \textbf{CR}$. Such matters are interesting and largely open problems that lie beyond the scope of this paper: we focus on $\textbf{CR}$ objects for concreteness.}

\subsection{Thermodynamic regularity}

{In contrast to the above}, the notion of thermodynamic regularity is less subtle because of the simple characterization of surface gravity (expression~\ref{eq:kapparot}), though care is still needed in discussing the heat capacity. For example, the entropy can disagree with the Bekenstein formula in different theories even for the same background metric \cite{voll07}: in an $f(R)$ theory, $S$ is weighted by $f'(R)$ and if $f'(0) \neq 1$ then $S \neq S^{\rm GR}$. We adopt the following definition:
\begin{itemize}
\item[]{\textbf{Thermodynamic regularity (TR):} $C_{\rm X}$ 
is bounded over phase space.}
\end{itemize}
In this sense, the \textbf{CR} condition can be checked by simply computing the Kretschmann invariant while \textbf{TR} can be checked by seeing if \eqref{eq:divcond} occurs anywhere. 

{In this work, we define ``doubly regular'' as an object that is both \textbf{CR} and \textbf{TR}. Subdivisions into different types of ``multi-regularity'' are possible however, including even further hierarchies relating to the (non-)existence of pseudospectral and other instabilities concerning BH perturbations \cite{dest21}. Future work will examine the set-theoretic relations between these notions.}

%%%%%%%%%%%%%%%%%%%%%%%%%%%%%%%%%%%%%%%%%%%%%%%%%%
\section{Static families} \label{sec:almostrn}
%%%%%%%%%%%%%%%%%%%%%%%%%%%%%%%%%%%%%%%%%%%%%%%%%%

We begin investigation of double regularity by considering a number of \textbf{CR} BHs. Many such objects have been constructed in the literature (especially those of a static variety), and we cannot reasonably consider all of them (especially because $\cal K$ is a complicated function of $\alpha$ and $B$). For practical purposes, it is convenient to choose families where the location of the horizon can be determined either analytically or through simple root-finding techniques, else computing $\kappa_{\cal H}$ and its thermodynamic derivatives becomes difficult. Typically, $\rH \equiv \rH(M,\boldsymbol{Q})$ and care must be exercised when differentiating while holding the ``hair'' $\boldsymbol{Q}$ (which need not represent charge but rather some generic hairs that may or may not have a non-Einstein origin), fixed. 

{Where needed, we use a brute-force method to locate roots. The function $\Delta(r)$ is tabulated with a small radial step ($\delta r = 10^{-6} M$) to find the limiting value of $r$ where a sign change occurs for a given value of $M$ (and other parameters/hairs). In practice, we find that using any value from $10^{-7} \lesssim \delta r/M \lesssim 10^{-5}$ does not change the results ($\ll 0.1\%$). This method is crude in the sense that it is inefficient relative to other root-finding routines (e.g. Newton-Raphson), but it suffices for our purposes. After finding roots for several nearby $M$, thermodynamic derivatives are taken numerically using a linear interpolation.}

%%%%%%%%%%%%%%%%%%%%%%%%%%%%%%%%%%%%%%%%%%%%%%%%%%
\subsection{Reissner-Nordstr{\"o}m metric} \label{sec:rnsec}
%%%%%%%%%%%%%%%%%%%%%%%%%%%%%%%%%%%%%%%%%%%%%%%%%%

Our starting point is the Schwarzschild solution  ($\alpha = 0, B = 1 - 2M/r$). The heat capacity in this case, at least in GR, is easily found as
\begin{equation} \label{eq:schheat}
C^{\rm Sch}_{\rm GR} = - 8 \pi M^2.
\end{equation}
Clearly, \eqref{eq:schheat} is non-positive over the entire phase space ($M \in \mathbb{R}^{+} \cup \{0\}$). In this sense, Schwarzschild objects are \textbf{TR} and ``cosmologically stable'' but not \textbf{CR} as $\mathcal{K} \propto r^{-6}$. 

Adding a nonzero charge, $Q$, yields the Reissner-Nordstr{\"o}m solution with $\alpha(r) = 0$ and 
\begin{equation} \label{eq:rn}
B(r) = 1 - \frac{2 M}{r} + \frac{Q^2}{r^2}.
\end{equation}
In this case, the heat capacity (in GR) is found as \cite{davies78}
\begin{equation} \label{eq:rnhc}
C^{\rm RN}_{\rm GR} = - \frac{2 \pi \sqrt{1-Q^2/M^2} \left(M + \sqrt{M^2 - Q^2}\right)^{2}}{2 \sqrt{1 - Q^2/M^2} -1}.
\end{equation}
It is easy to see that the denominator of expression \eqref{eq:rnhc} diverges as $Q \to Q^{\rm RN}_{\rm D} = \pm\sqrt{3}M/2$, meaning that a Davies' point exists where
\begin{equation} \label{eq:daviesRN}
\lim_{|Q| \to \sqrt{3} M /2} C^{\rm RN}_{\rm GR} \to \infty.
\end{equation}
While astrophysical charge is always expected to be below the limiting value ($|Q| \ll Q^{\rm RN}_{\rm D}$), it is clear that the RN spacetime is neither \textbf{CR} nor \textbf{TR}.

%%%%%%%%%%%%%%%%%%%%%%%%%%%%%%%%%%%%%%%%%%%%%%%%%%%%%%%%%%%%
\subsection{Simpson-Visser type} \label{sec:svtype}
%%%%%%%%%%%%%%%%%%%%%%%%%%%%%%%%%%%%%%%%%%%%%%%%%%%%%%%%%%%%

Consider the (Minkowski-core) Simpson-Visser (SV) \cite{sv19} metric\footnote{Note there are several metrics sharing this name; the one used here corresponds to expression (37) in Ref.~\cite{sv19}.} with a free parameter $k$, viz. $\alpha(r) = 0$ and 
\begin{equation} \label{eq:sv}
B(r) = 1 - \frac{2M}{r} e^{-Q^{k}/r^{k}}.
\end{equation}
Using the potential \eqref{eq:sv}, one finds that
\begin{equation}
\begin{aligned}
\mathcal{K}^{\rm SV} =\,& \frac{4 M^2 e^{-2 Q^k/ r^{k}}}{r^{4 k+6}} \Big[k^4 Q^{4 k}-2 k^3 \left(k+3\right) Q^{3 k} r^k \\
&+ k^2 \left(k^2+6k+17\right) Q^{2 k} r^{2 k}\\
&-4 k \left(k+5\right) Q^k r^{3 k}+12 r^{4 k}\Big],
\end{aligned}
\end{equation}
and, as such, $\mathcal{K}^{\rm SV}$ is everywhere bounded and well-behaved for $Q>0$ and integer $k$: the spacetime is \textbf{CR}.

\begin{figure}
\begin{center}
\includegraphics[width=0.487\textwidth]{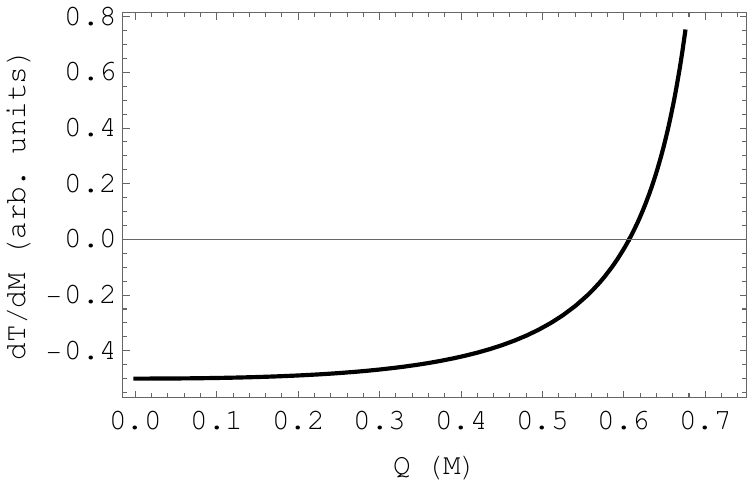}
\caption{Variation of the horizon temperature as a function of general hair, $Q$, for the SV family \eqref{eq:sv} with $k=1$. A Davies' point is visible at $Q = M/\sqrt{e}$.
\label{fig:sv}}
\end{center}
\end{figure}

The horizon is found via the principal branch of the Lambert W function,
\begin{equation}
\rH(M,Q) = \left[- \frac{k Q^{k}}{W_{0}\left(-k Q^k/2^k M^k\right)}\right]^{1/k},
\end{equation}
so that computing $\kappa_{\cal H}$ is relatively straightforward, though the expression is lengthy. Considering the case of $k=1$ for demonstration, it is easy to see that an event horizon exists provided $Q \leq 2 M /e$; the horizon radius is $\rH = 2M/e$ for this extremal case and we find
\begin{equation}
\left(\frac{\partial \kappa_{\cal H}}{\partial M}\right)_{Q} = \frac{W_{0}(-Q/2M) \left[ 1 + 2 W_{0}(-Q/2M) \right]}{M Q \left[ 1 + W_{0}(-Q/2M)\right]},
\end{equation}
which has a zero at $Q_{\rm D} = M/\sqrt{e} < 2M/e$. Figure~\ref{fig:sv} depicts the relationship $(\partial T/\partial M)_{Q}$ as a function of $Q$, clearly showing the Davies' point. Cases with $k > 1$ also suffer from thermodynamic singularities. They are, therefore, not \textbf{TR} despite a number of other attractive features \cite{sv19}. Note that prior to the Davies' point the heat capacity is \emph{negative}, as for Schwarzschild \eqref{eq:schheat}, indicating ``stability'' in the sense described in Sec.~\ref{sec:physical}.

%%%%%%%%%%%%%%%%%%%%%%%%%%%%%%%%%%%%%%%%%%%%%%%%%%%%%%%%%%%%
\subsection{Weighted Hayward holes} \label{sec:whtype}
%%%%%%%%%%%%%%%%%%%%%%%%%%%%%%%%%%%%%%%%%%%%%%%%%%%%%%%%%%%%

Another popular set of regular BH metrics are the Hayward family \cite{hay06}. We consider a simple generalization for reasons that will become clear -- a ``weighted Hayward'' metric -- with 
\begin{equation} \label{eq:hay1}
\alpha(r) = \log\left[ 1 - \frac{\zeta M}{\left(Q - r \right)} \right],
\end{equation}
and 
\begin{equation} \label{eq:hay2}
B(r) = 1 - \frac{2 M r^k}{M Q^{k} + r^{k+1}},
\end{equation}
for fixed dimensionless parameters $k \geq 1$ and $\zeta$ with hair $Q$. Setting $\zeta = 0$ returns Hayward for $k=2$: this case is similar to SV in that the metrics are \textbf{CR} but not \textbf{TR} as a Davies' point appears at $Q = Q_{\rm D} \approx 0.84 M$ \cite{mol21}. The fact that $Q_{\rm D}$ is large is enticing as it suggests that, perhaps with only small tweaks, a generalized metric may be \textbf{TR}.

\begin{figure}
\includegraphics[width=0.487\textwidth]{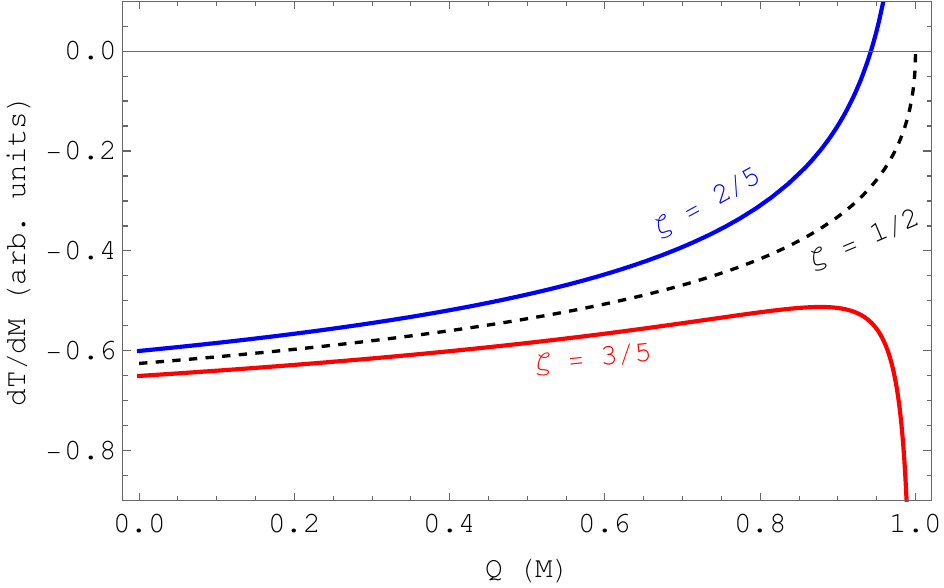}
\caption{Similar to Fig.~\ref{fig:sv} but for the weighted Hayward family (defined by expressions \ref{eq:hay1} and \ref{eq:hay2}) with $k=1$ for $\zeta = 2/5$ (solid blue), $\zeta = 1/2$ (dashed black), or $\zeta = 3/5$ (red). 
\label{fig:hayward}}
\end{figure}

At least for the case $k=1$ and $\zeta > 1/2$, inspection of $\kappa_{\cal H}$ reveals that there are no Davies' points: they migrate outward\footnote{For $\zeta < 0$ the value of $Q_{\rm D}$ decreases below that of pure Hayward and, although not shown, for $\zeta < -2$ there are also no Davies' points with the heat capacity instead being \emph{positive} for $Q >0$.} (from a phase-space perspective) for positive $\zeta$, pushing past extremality for $\zeta > 1/2$. This is depicted in Figure~\ref{fig:hayward}, showing $dT/dM$ as a function of $Q$ for $k=1$ up to the extremal limit ($Q=M$) for a few values of $\zeta$. For the $\zeta = 2/5$ case (blue), a Davies' point is visible for $Q_{\rm D} \approx 0.94$ which, while larger than for pure Hayward as described above, still resides within the permissible phase space ($Q \leq M$). For the critical value $\zeta = 1/2$ (dashed), we see that this migration pushes $Q_{\rm D}$ to unity, indicating that geometric and phase extremality are met at the same ``charge''. When $\zeta > 1/2$ (red), however, there is a turnover behavior such that $dT/dM < 0$ for all $Q \leq M$, meaning that the object is \textbf{TR}. Unfortunately though, this property does not persist for larger $k$ as increasing $\zeta$, while still pushing $Q_{\rm D}$ outward, keeps the Davies' point below the extremal limit (i.e., to values $Q>M$). The metric is therefore \textbf{TR} only for $k=1$.

On the other hand, a routine calculation reveals that
\begin{equation} \label{eq:kretk1}
\mathcal{K}^{\rm WH}(k=1) \underset{r \to 0}{\sim} \frac{8}{r^2} \left[ \frac{3}{Q^2} + \frac{1}{\left(Q - M \zeta \right)^{2}} \right],
\end{equation}
which cannot be made to vanish for any real-valued choice of $\zeta$. This is in contrast to the $k=2$ case where
\begin{equation} \label{eq:kretk2}
\mathcal{K}^{\rm WH}(k=2) \underset{r \to 0}{\sim} \frac{8 M^2 \zeta^2}{Q^2 \left(Q - M \zeta \right)^2 r^2},
\end{equation}
for which the divergent behavior is neutralized if $\zeta = 0$ (i.e., pure Hayward). As such, while weighted Hayward objects can flip between \textbf{CR} and \textbf{TR} states they cannot be both simultaneously for any parameter choices.

%%%%%%%%%%%%%%%%%%%%%%%%%%%%%%%%%%%%%%%%%%%%%%%%%%%%%%%%%%%%%%%%%%%%%%%%%%%
\subsection{Hybrid Bardeen type} \label{sec:bardeen}
%%%%%%%%%%%%%%%%%%%%%%%%%%%%%%%%%%%%%%%%%%%%%%%%%%%%%%%%%%%%%%%%%%%%%%%%%%%

While the Bardeen metric itself is known to have a Davies' point \cite{myung07}, as are the extended class introduced in Ref.~\cite{fw16}, we construct here a generalization that holds promise for double regularity. Consider $\alpha(r) = 0$ and 
\begin{equation} \label{eq:bardeen}
B(r) = 1 - \frac{2M}{r}e^{-\zeta M/r} - \frac{M r^{k-1}}{\left(r + Q \right)^{k}},
\end{equation} 
which resembles a hybrid between SV and pure Bardeen with a dimensionless constant $\zeta$. Importantly, the third term promotes thermodynamic regularity as it generally works to \emph{reduce} the heat capacity for positive $Q$, which may temper the SV Davies' point (see Fig.~\ref{fig:sv}). Note that an event horizon exists for any $Q$ for suitably small $\zeta$.

\begin{figure}
\includegraphics[width=0.487\textwidth]{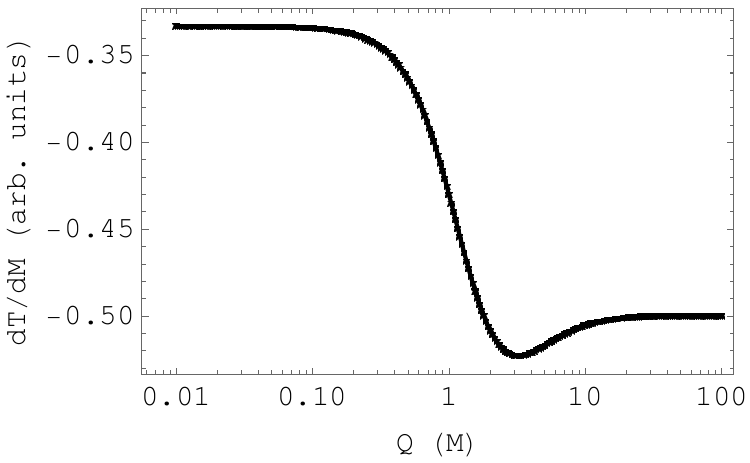}
\caption{Similar to Fig.~\ref{fig:sv} but for a Bardeen-like BH (expression \ref{eq:bardeen}) with $k=3$ and $\zeta = 1/100$. Numerically-computed values of $(\partial T/\partial M)_{Q}$ are shown by stars with the solid curve rendering a linear interpolation.
\label{fig:bardeen}}
\end{figure}

We find that, for $k \geq 3$ and small $\zeta$ ($\ll 1$), these objects are that which we seek: they are both \textbf{CR} (as is clear from the exponential suppression of the monopole term and the cubic, Bardeen-like nature of the third term) and \textbf{TR}. The latter property is demonstrated in Fig.~\ref{fig:bardeen} depicting the variation of temperature with respect to mass (for fixed constant $\zeta=1/100$ and varying hair $Q$), which has to be computed numerically {(see the discussion at the start of Sec.~\ref{sec:almostrn})} owing to the transcendental nature of the potential \eqref{eq:bardeen} We see that $\partial T/\partial M$ plateaus after $Q \gtrsim 50$, and moreover is negative everywhere. 

At least within the restricted context of static BHs, of all those considered thus far only the hybrid Bardeen-like family \eqref{eq:bardeen} is \emph{doubly regular}. Although we have clearly not provided an exhaustive survey of regular objects, the study of these families is sufficient to demonstrate that double regularity is a rather restrictive condition. While a more involved study of astrophysical manifestations -- if any -- is required to assess whether this mathematically-privileged position is of physical relevance (e.g. with respect to radiative balding \cite{page76}), the explicit construction of the doubly regular metric \eqref{eq:bardeen} forms one of our main results in this work. {Some brief discussion on astrophysical scenarios is given in Sec.~\ref{sec:discussion}.}

%%%%%%%%%%%%%%%%%%%%%%%%%%%%%%%%%%%%%%%%%%%%%%%%%%
\section{Rotating extensions} \label{sec:rotfams}
%%%%%%%%%%%%%%%%%%%%%%%%%%%%%%%%%%%%%%%%%%%%%%%%%%

Although the Newman-Janis algorithm tends to preserve the \textbf{CR} property of BH spacetimes \cite{bamb13}, given that inserting $a=0$ into the metric \eqref{eq:njmet} returns the static case it is clear that the inclusion of rotation cannot, by itself, induce regularity. Moreover, angular momentum works to increase the heat capacity in the Kerr case \cite{davies78}. This means that in the limit of small hair but large angular momentum where Kerr features dominate, it appears that rotating \textbf{TR} metrics are difficult to construct. 

As far as double regularity is concerned, we only need to study the Bardeen-like family considered in Sec.~\ref{sec:bardeen} as the static versions of others do not share this property.

%%%%%%%%%%%%%%%%%%%%%%%%%%%%%%%%%%%%%%%%%%%%%%%%%%%%%%%%%%%%
\subsection{Rotating Bardeen-like black hole} \label{sec:rotbard}
%%%%%%%%%%%%%%%%%%%%%%%%%%%%%%%%%%%%%%%%%%%%%%%%%%%%%%%%%%%%

As depicted in Fig.~\ref{fig:bardeen}, the static, hybrid Bardeen family \eqref{eq:bardeen} can be made both \textbf{TR} and \textbf{CR} for suitable parameter choices. Unfortunately however, the inclusion of rotation introduces Davies' points irrespective of the value of $Q$ for this class, much like in the Schwarzschild case. 

\begin{figure}
\begin{center}
\includegraphics[width=0.487\textwidth]{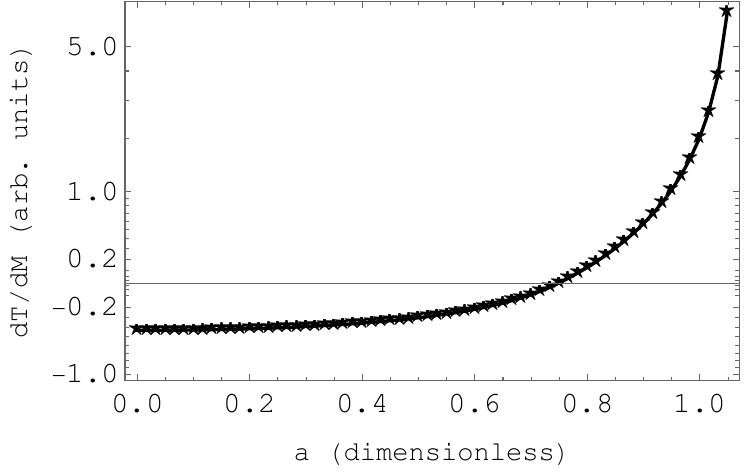}
\caption{Variation of the horizon temperature with mass as a function of spin for the rotating Bardeen-like metric (expression \ref{eq:njmet} seeded by \ref{eq:bardeen}) with $Q=1$ and $\zeta = 1/100$.
\label{fig:bard1}}
\end{center}
\end{figure}

Figure~\ref{fig:bard1} shows temperature variations using equation \eqref{eq:kapparot} for the metric defined by equation \eqref{eq:njmet} for the seed \eqref{eq:bardeen} with $k=3$, $Q=1$, and $\zeta = 1/100$ as a function of spin (as in Fig.~\ref{fig:bardeen}). A numerical root-finder shows that a horizon exists provided that $a \lesssim 1.0653$. Although a positive $Q$ generally helps to thermodynamically stabilize the object (Fig.~\ref{fig:bardeen}), we see that $\partial T / \partial M$ changes sign around $a \approx 0.75$. This demonstrates that simply taking $Q,\zeta > 0$ is insufficient to make the object \textbf{TR} (for any $k$); fixing different values of $\zeta$ does not remedy the problem. 

Although a different rotational extension might be able to mitigate this behavior (see Sec.~\ref{sec:rotating}), none of the \textbf{CR} objects investigated thus far are \textbf{TR} when considering axially-symmetric extensions. %In fact, we are unaware of any such example from the literature.

%%%%%%%%%%%%%%%%%%%%%%%%%%%%%%%%%%%%%%%%%%%%%%%%%%
\subsection{Proof of concept: stationary CFLV metric} \label{sec:hosv}
%%%%%%%%%%%%%%%%%%%%%%%%%%%%%%%%%%%%%%%%%%%%%%%%%%

The final set of results in this paper concerns the explicit construction of a stationary, \textbf{TR} object. Consider the metric function
\begin{equation} \label{eq:higherorder}
B(r) = 1 - \frac{2 M \left(r^2 + \zeta_{1}^2\right)}{\left(r^2+ \zeta_{1}^2 \right)^{3/2} + 2 M \zeta_{2}^2},
\end{equation}
for some parameters $\zeta_{1,2}$ which have units of length. In particular, we suppose that these parameters are fixed by \emph{dimensionful} constants appearing within some hitherto unknown gravitational action (resembling, e.g., Newton's constant) rather than emerging as hairs which can vary from object to object in the sense of boundary data applied to the equations of motion (resembling, e.g., electric charge). We refer to metrics defined by the potential \eqref{eq:higherorder} as being ``CFLV'' after the names of the authors behind Chapter 9.2 of Ref.~\cite{bamb23}. The metric has both Simpson-Visser (in the sense of Ref.~\cite{svalt}; cf. Footnote 2) and Hayward limits for $\zeta_{1} \to 0$ or $\zeta_{2} \to 0$, respectively. It could thus also be referred to as a hybrid Hayward-Simpson-Visser object. A discussion of the geometric properties of this spacetime can be found in Chapter 9.2 of Ref.~\cite{bamb23}, though to our knowledge rotating generalizations have not been considered in the literature before. We have checked that the rotating variant is \textbf{CR} \cite{bamb13}. 

One key property of the metric \eqref{eq:higherorder} is that $\zeta_{1}$ works to increase the heat capacity in general, much like the spin parameter. This is important because, unlike the Schwarzschild case ($\zeta_{1,2}=0$), one may choose $\zeta_{1}$ sufficiently large (for given $M$) such that $dT/dM$ is positive in the static limit. As such, switching on rotation -- which tends to increase it further -- preserves a \textbf{TR} spacetime.  

\begin{figure}
\includegraphics[width=0.487\textwidth]{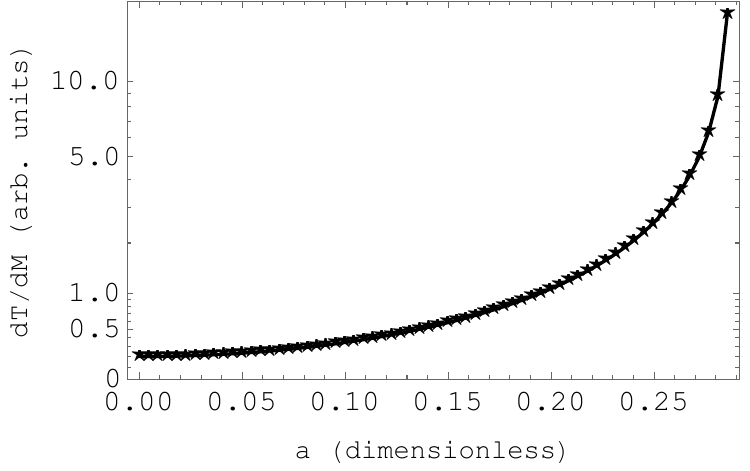}
\caption{Similar to Fig.~\ref{fig:bard1} but for the metric potential \eqref{eq:higherorder}, for fixed parameter choices $M=1$, $\zeta_{1} = 3/2$, and $\zeta_{2} = 1/\sqrt{10}$.
\label{fig:gensch}}
\end{figure}

\begin{table*}
	\caption{Summary of the key properties and results for the various metric families considered in this work. \textbf{Notes}. $\ast$ For $\zeta = 0$ we have \textbf{CR} not \textbf{TR}. $\dagger$ Since the parameter $\zeta$ is dimensionful, the \textbf{TR} condition may depend on $M$.
}
\begin{tabular}{l|cc|cc|c|c}
	    \hline
	    Metric family & $\alpha(r)$ & $B(r)$ & \textbf{CR} ($a \neq 0$)? & \textbf{TR} ($a \neq 0$)? & Conditions & Reference \\
	    	    \hline
	    	    	    \hline
Schwarzschild & $0$ & $1 - \frac{2M}{r}$ &  \xmark (\xmark) & \cmark (\xmark) & -- & \cite{davies78} \\    
Reissner-Nordstr{\"o}m & $0$ & $1 - \frac{2M}{r} + \frac{Q^2}{r^2}$ &  \xmark (\xmark) & \xmark (\xmark) & -- & \cite{davies78} \\
Extended Simpson-Visser & $0$ & $1 - \frac{2M}{r} e^{-Q^{k}/r^{k}}$ & \cmark (\cmark) & \xmark (\xmark) & $k \in \mathbb{N}, Q>0$ & \cite{sv19} \\
Weighted Hayward & $\log\left[ 1 - \frac{\zeta M}{\left(Q - r \right)} \right]$ & $1 - \frac{2 M r^{k}}{M Q^{k} + r^{k+1}}$ & \xmark* & \cmark* & $\zeta > 1/2$, $k \geq 1$ & This work \\
Hybrid Bardeen-like & $0$ & $1 - \frac{2M}{r}e^{-\zeta M/r} - \frac{ Q r^{k-1}}{\left(r + Q \right)^{k}}$ & \cmark (\cmark) & \cmark (\xmark) & $\zeta \ll 1, k \geq 3$ & This work \\
Stationary CFLV & $0$ & $1 - \frac{2 M \left(r^2 + \zeta_{1}^2\right)}{\left(r^2+ \zeta_{1}^2 \right)^{3/2} + 2 M \zeta_{2}^2} $ & \cmark (\cmark) & \cmark (\cmark$^{\dagger}$) & $\zeta_{1} \gtrsim M, \zeta_{2} \neq 0$ & \cite{bamb23} \\
     \hline
	    \hline
	\end{tabular}
	\label{tab:keyprops}
\end{table*}

Figure~\ref{fig:gensch} shows the variation of temperature for $M=1$ as a function of dimensionless spin for $\zeta_{1} = 3/2$ and $\zeta_{2} = 1/\sqrt{10}$. For these choices, a routine root-finding demonstrates that a real solution to $\Delta(\rH) = 0$ exists for $a \lesssim 0.2901$. We see that the capacity is always positive which, although again potentially indicating a cosmologically-questionable scenario (see Sec.~\ref{sec:physical}), illustrates the absence of Davies' points. It is therefore, in principle, possible for stationary spacetimes with an obvious Kerr limit to be both \textbf{CR} and \textbf{TR}. This may not hold for arbitrary $M$ though, as the \textbf{TR} property is sensitive to the exact value of $\zeta_{1}$ and it must be chosen to be commensurately large. If there was a maximum mass limit for a physical BH (e.g. set by the energy budget of the universe), then one could guarantee that no such object could reach a state that is not doubly regular. While the set of axially-symmetric and asymptotically-flat doubly-regular objects is therefore not empty, it does appear especially restrictive. This new example at least provides a simple proof-of-concept: rotating, doubly regular black holes can exist. How large such a set is, or whether cases with everywhere negative capacities exist, remain open questions however reserved for future work. 

An executive summary of the key properties of metrics considered in this work is provided in Table~\ref{tab:keyprops}.

%%%%%%%%%%%%%%%%%%%%%%%%%%%%%%%%%%%%%%%%%%%%%%%%%%
\section{Discussion} \label{sec:discussion}
%%%%%%%%%%%%%%%%%%%%%%%%%%%%%%%%%%%%%%%%%%%%%%%%%%

While geometric singularities are indicative of high-energy incompleteness and have been studied extensively, singularities of a Davies' nature also warrant attention owing to deep connections between thermodynamics and gravity \cite{davies78,wald93,jacob95,pad02}. Many curvature-regular BHs have been introduced in the literature \cite{torres23,lan23} -- that called \textbf{CR} in this paper -- and often appear practically indistinguishable from their GR counterparts on an astrophysical level because regulators are tied to Planck-scale physics. We have shown that those free of Davies' points -- thermodynamically-regular or \textbf{TR} spacetimes -- appear to form a very restrictive subset of regular BHs. Of those static \textbf{CR} BHs considered in Sec.~\ref{sec:static}, we saw that \emph{none} of them were simultaneously \textbf{TR} (i.e. doubly regular) except for a new hybrid mixture of the SV and Bardeen families (expression \ref{eq:bardeen}), demonstrating that it is possible for a physically-motivated object to satisfy this property. 

If including dimensionful constants within the metric that are not to be considered hairs but rather fixed parameters within a given theory, we demonstrated in Sec.~\ref{sec:hosv} that it is also possible to construct axially-symmetric objects that are doubly regular. This mathematical property appears to be very restrictive though, and there are currently no known examples of other such spacetimes in the literature that we are aware of (appearing as exact solutions in a particular theory or otherwise). Although the simple example we constructed was shown to be free of Davies' points for sufficiently large and fixed $\zeta_{1,2}$ (Fig.~\ref{fig:gensch}), it is not especially satisfying because the constant is dimensionful and large. Future work will be devoted to searching for doubly-regular geometries, including other geometric ingredients such as spacetime asymptotics or topological elements of the BH interior (see Ref.~\cite{barg20}). Such aspects may invite further subdivisions into notions of multiply-regular holes, {as discussed in Sec.~\ref{sec:notions}.} 

{In terms of astrophysical manifestations, there are a few paths that one could imagine to distinguish a doubly regular object from a singularly regular BH or otherwise. \citet{bolo24} has shown that massive-field perturbations of the Bardeen BH lead to quasi-normal modes with $\sim 1/t$ falloffs at late times, which differs from Schwarzschild perturbations. These modes (``quasi-resonances'') can be long-lived and thus, in principle, could be resolved with some (far-future) detector. If the character of these modes change for the hybrid Bardeen metric \eqref{eq:bardeen}, this may provide an astrophysical tool to constrain double regularity (i.e., by comparing data and predictions for the Schwarzschild vs. Bardeen vs. hybrid Bardeen cases). Another aspect highlighted in Sec.~\ref{sec:physical} relates to the radiative power surrounding heavy objects. While a detection of the spectral properties of quantum-mechanical radiation from BHs is unlikely anytime soon, analogue gravity experiments could provide insight into the nature of thermodynamic phase transitions \cite{silke13}.}

The dynamical stability of doubly-regular objects would also be interesting to consider, in the sense of whether perturbations can grow in amplitude with time or if the perturbed Hawking temperature \eqref{eq:temp} adjusts such that a Davies' point may emerge. The methods introduced in Refs.~\cite{suvv21,suvx21} could be used, for instance, to construct theories around a given doubly-regular spacetime to study its gravitational perturbations.

\section*{Acknowledgements}

Support from the Conselleria d'Educaci{\'o}, Cultura, Universitats i Ocupaci{\'o} de la Generalitat Valenciana thorugh Prometeo Project CIPROM/2022/13 is gratefully acknowledged.

%

%%%%%%%%%%%%%%%%%%%%%

\end{document}